\documentclass[aps,prc,twocolumn,superscriptaddress,10pt]{revtex4-2}

\usepackage{graphicx}
\usepackage{dcolumn}
\usepackage[version=4]{mhchem}
\usepackage{amsmath}
\usepackage{amssymb}

\usepackage[greek,russian,spanish,english]{babel}
\usepackage{url}
\usepackage[colorlinks,
linkcolor=blue,
anchorcolor=blue,
urlcolor=blue,
citecolor=blue]{hyperref}
\begin{document}


\title{ Particle number projected energies at finite temperature}
\author{Jiawei Chen}
\affiliation{State Key Laboratory of Nuclear Physics and Technology, School of Physics,
Peking University, Beijing 100871, China}

\author{Yu Qiang}
\affiliation{State Key Laboratory of Nuclear Physics and Technology, School of Physics,
Peking University, Beijing 100871, China}

\author{Junchen Pei}
\email{peij@pku.edu.cn}
\affiliation{State Key Laboratory of Nuclear Physics and Technology, School of Physics,
Peking University, Beijing 100871, China}

\date{\today}

\begin{abstract}
In this work, the particle number projection at finite temperature is incorporated into self-consistent Skyrme density functional calculations.
In particular, the energies of compound nuclei as a function of deformations are calculated rigorously based on projected densities.
Results show that the even-odd staggering effect in partition function gradually diminishes as the system approaches the critical temperature.
The obtained fission barriers are similar to that  without projection at finite temperature, although projected energies are different.
The nuclear level density at the ground state and the barrier are also studied using the projection method and the discrete Gaussian method.
\end{abstract}

\maketitle

\section{Introduction}

Nuclear energy density functional theory (DFT) has advantages in descriptions of nuclear properties and dynamics of heavy and superheavy nuclei\cite{benderSelfconsistentMeanfieldModels2003}.
In particular, DFT can play an essential role in addressing theoretical issues regarding the synthesis of superheavy nuclei\cite{qiangSurvivalProbabilitiesCompound2024}, nuclear fission and fusion reactions\cite{scampsImpactPearshapedFission2018,bulgacTimeDependentDensityFunctional2025,qiangQuantumEntanglementNuclear2025}, and large-scale calculations of nuclear masses\cite{stoitsovLargescaleMassTable2009,guanHighQualityMicroscopic2024}, density distributions\cite{zhouNeutronHaloDeformed2010,pei2005density}, and potential energy surfaces\cite{wardaFissionHalflivesSuperheavy2012,marevicFissionPu2402020,nSchunck2014PotentialEnergySurface,chi2023role}.
For heavy and superheavy nuclei, the strong repulsive Coulomb potential can change density distributions and shell structures so that
self-consistent DFT calculations are needed\cite{nazarewiczTheoreticalDescriptionSuperheavy2002,pei2005density}. In addition, the temperature dependence (or energy dependence) can be self-consistently included
in the DFT framework\cite{alanl.goodmanFinitetemperatureHFBTheory1981,peiFissionBarriersCompound2009}.
However, DFT is associated with the symmetry breaking at the mean-field level\cite{sheikhSymmetryRestorationMeanfield2021,ringNuclearManyBody2004}. To this end, various projection
methods such as the particle number projection (PNP) and angular momentum projection to restore the broken symmetries have been developed\cite{m.v.stoitsovVariationParticlenumberProjection2007,bertschSymmetryRestorationHartreeFockBogoliubov2012,bulgacProjectionGoodQuantum2019,verriereMicroscopicCalculationFission2021,ballyProjectionParticleNumber2021}.

The pairing correlations as an important ingredient of DFT are conventionally described by the
BCS theory or the Bogoliubov transformation\cite{ringNuclearManyBody2004}. In these approaches, the gauge symmetry is broken, which leads to the mixing of configurations with different particle numbers.
Consequently the number of particles is not a good quantum number owing to the particle number fluctuations, although
the expectation value of particle number can be constrained\cite{sheikhSymmetryRestorationMeanfield2021}.
Furthermore, the nuclear thermal excitations are usually described in the grand canonical ensemble,  which inherently introduces
particle number fluctuations.
 In the finite temperature Hatree-Fock+BCS or Hartree-Fock-Bogoliubov approaches\cite{alanl.goodmanFinitetemperatureHFBTheory1981},
the particle number fluctuations from both the grand canonical ensemble and the gauge symmetry breaking in BCS theory coexist\cite{rossignoliProjectionFiniteTemperature1994}.
Eventually the pairing and quantum effects would be washed out as the temperature increases.
The pairing at finite temperature is also an interesting subject for quantum computing\cite{jiangQuantumComputingPairing2023}. 
The non-conservation of particle numbers is a non-negligible issue in finite nuclei.
The saddle point approximation and discrete Gaussian approximations(DG) have been used to restore the particle numbers in the grand canonical ensemble\cite{y.alhassidBenchmarkingMeanfieldApproximations2016}.
For superfluid systems, the restoration of particle number conservation can be achieved through higher-order constrains on particle numbers using
the  Lipkin-Nogami method\cite{nogamiImprovedSuperconductivityApproximation1964,wang2014Lipkin}.
The exact particle number projection can not only restore the particle numbers but also calculate projected observables.
The PNP formulism at finite temperature would be more complicated compared to PNP at zero temperature.
At zero temperature, the projection operator is directly applied to the ground-state wave function\cite{sheikhSymmetryprojectedHartreeFock2000,anguianoParticleNumberProjection2001,m.v.stoitsovVariationParticlenumberProjection2007}. In contrast, the projection operator applied to the ensemble at finite temperature\cite{rossignoliFiniteTemperatureProjected1992,esebbagNumberProjectedStatistics1993,rossignoliProjectedStatisticsLevel1993,rossignoliProjectionFiniteTemperature1994}.

The PNP at finite temperature has been developed by Fanto et al. to calculate the projected partition function\cite{fantoParticlenumberProjectionFinitetemperature2017,p.fantoProjectionVariationFinitetemperature2017}.
Based on the partition function, the entropy and level density of compound nuclei can be obtained.
The approximate excitation energy can also be obtained by the partial derivatives of the partition function.
The aim of this work is to calculate  PNP energies at finite temperature exactly based on the projected ensemble and densities.
This is relevant to energy dependent fission barriers and excitation energies of fission fragments.
Note that PNP energies at zero temperature has been calculated exactly\cite{anguianoParticleNumberProjection2001}.
However, such calculations at finite temperature
are more difficult and have not been realized yet.
Furthermore, PNP at finite temperature has been testified with the Monte Carlo shell model\cite{fantoParticlenumberProjectionFinitetemperature2017} but it has never been implemented
in self-consistent DFT calculations.

In this work, the formulism to calculate PNP energies at finite temperature are derived,
based on the densities in terms of creation and annihilation operators.
The formulism is applied to calculate the exact PNP energies of compound nuclei.
Presently PNP after variation is implemented. 
In principle, the variation after projection  is more accurate since the projection after variation violates the variational principle\cite{zehSymmetryViolatingTrial1965},
but the variation after projection is very complicated. Moreover, in self-consistent calculations by minimizing the free energy,
it is practically impossible to apply Wick's theorem to the matrix elements of the entropy, which prohibits the variation after projection at finite temperature\cite{esebbagNumberProjectedStatistics1993}.
 Based on finite-temperature Skyrme-Hartree-Fock+BCS calculations, we investigate the fission barriers of heavy and superheavy nuclei  $\ce{^{238}U}$ and $\ce{^{292}Fl}$
 with and without PNP.
 We also studied the level densities using the PNP partition function and the DG method.
 Furthermore, the level density paramters at the ground state deformation and the barrier are extracted, which are useful
 for statical models to calculate survival probabilities of compound superheavy nuclei\cite{qiangSurvivalProbabilitiesCompound2024}.

\section{Methods}

Considering a quantum system described by a density matrix $\hat{D}$ in the grand canonical ensemble, the corresponding system with fixed particle number can be described by a projected density matrix \cite{rossignoliProjectionFiniteTemperature1994}

\begin{equation}
    \hat{D}_{\text{p}} = \frac{1}{Z_{\text{p}}}\hat{P}\hat{D}\hat{P},
\end{equation}
where the partition function $Z_{\text{p}}$ is given by $\mathop{\text{Tr}}(\hat{P}\hat{D})$, as described in \cite{esebbagNumberProjectedStatistics1993, p.fantoProjectionVariationFinitetemperature2017, robledoSignOverlapHartreeFockBogoliubov2009}. The particle number projection operator $\hat{P}$, which projects out a state with particle number $N$, is expressed as a standard integral form in terms of the gauge angle $\theta$ as:

\begin{equation}
    \hat{P}=\frac{1}{2\mathrm{\pi}}\int\mathrm{d}\theta\,\mathrm{e}^{\mathrm{i}\theta(\hat{N}-N)}.
\end{equation}

The expectation value of a physical quantity $\hat{O}$ is given by $\mathop{\text{Tr}}(\hat{D}_{\text{p}}\hat{O})$.
Considering the commutativity between $\hat{P}$ and  $\hat{O}$, the expression can
be simplified as $\langle\hat{O}\rangle=\mathop{\text{Tr}}(\hat{P}\hat{D}\hat{O})/Z_{\text{p}}$.
 Consequently, the energy of the system is given by:

\begin{equation}
    E=\frac{1}{Z_{\text{p}}}\frac{1}{2\mathrm{\pi}}\int\mathrm{d}\theta\,\mathrm{e}^{-\mathrm{i}\theta N}\mathop{\text{Tr}}(\mathrm{e}^{\mathrm{i}\theta\hat{N}}\hat{D}\hat{H}).\label{energy}
\end{equation}

By defining $\mathop{\text{Tr}}(\mathrm{e}^{\mathrm{i}\theta\hat{N}}\hat{D}\hat{H})=\mathop{\text{Tr}}(\mathrm{e}^{\mathrm{i}\theta\hat{N}}D)E(\theta)$, the quantity $E(\theta)$ can be constructed in terms of single-particle density matrices $\tilde{\rho}, \tilde{\lambda}$ and $\tilde{\kappa}$ using the generalized Wick theorem\cite{tanabeExtensionWickTheorem1999}, yielding:

\begin{align}
    E  (\theta)&=\sum_{pq}\langle p|f|q\rangle\tilde{\rho}_{qp}+\frac{1}{2}\sum_{pqrs}\langle pq|\bar{v}|sr\rangle\tilde{\rho}_{sp}\tilde{\rho}_{rq}\nonumber\\
    &+\frac{1}{4}\sum_{pqrs}\langle pq|\bar{v}|sr\rangle\tilde{\lambda}_{pq}\tilde{\kappa}_{sr},\label{gauged_energy}
\end{align}

where $\langle pq|\bar{v}|rs\rangle=\langle pq|v|rs\rangle-\langle pq|v|sr\rangle$, and the single-particle density matrices $\tilde{\rho}$, $\tilde{\lambda}$ and $\tilde{\kappa}$ are defined as

\begin{align}
    \tilde{\rho}_{qp}&=\frac{\text{Tr}(\mathrm{e}^{\mathrm{i}\theta\hat{N}}\hat{D}a^{\dagger}_{p}a_{q})}{\text{Tr}(\mathrm{e}^{\mathrm{i}\theta\hat{N}}\hat{D})},\\
    \tilde{\kappa}_{qp}&=\frac{\text{Tr}(\mathrm{e}^{\mathrm{i}\theta\hat{N}}\hat{D}a_{p}a_{q})}{\text{Tr}(\mathrm{e}^{\mathrm{i}\theta\hat{N}}\hat{D})},\\
    \tilde{\lambda}_{pq}&=\frac{\text{Tr}(\mathrm{e}^{\mathrm{i}\theta\hat{N}}\hat{D}a^{\dagger}_{p}a^{\dagger}_{q})}{\text{Tr}(\mathrm{e}^{\mathrm{i}\theta\hat{N}}\hat{D})}.
\end{align}

The detailed expression of these three single-particle density matrices can be written as,
\begin{align}
\tilde{\rho}_{qp}=\delta_{pq}\frac{\mathrm{e}^{-2\beta E_{p}}\mathrm{e}^{2\mathrm{i}\theta}u_{p}^{2}+\mathrm{e}^{-\beta E_{p}}\mathrm{e}^{\mathrm{i}\theta}+\mathrm{e}^{2\mathrm{i}\theta}v_{p}^{2}}{u_{p}^{2}+\mathrm{e}^{2\mathrm{i}\theta}v_{p}^{2}+2\mathrm{e}^{-\beta E_{p}}\mathrm{e}^{\mathrm{i}\theta}+\mathrm{e}^{-2\beta E_{p}}(\mathrm{e}^{2\mathrm{i}\theta}u_{p}^{2}+v_{p}^{2})},\\
\tilde{\kappa}_{qp}=\delta_{p\bar{q}}\frac{\mathrm{e}^{2\mathrm{i}\theta}(\mathrm{e}^{-2\beta E_{p}}u_{p}v_{p}-u_{p}v_{p})}{u_{p}^{2}+\mathrm{e}^{2\mathrm{i}\theta}v_{p}^{2}+2\mathrm{e}^{-\beta E_{p}}\mathrm{e}^{\mathrm{i}\theta}+\mathrm{e}^{-2\beta E_{p}}(\mathrm{e}^{2\mathrm{i}\theta}u_{p}^{2}+v_{p}^{2})},\\
\tilde{\lambda}_{pq}=\delta_{p\bar{q}}\frac{-\mathrm{e}^{-2\beta E_{p}}u_{p}v_{p}+u_{p}v_{p}}{u_{p}^{2}+\mathrm{e}^{2\mathrm{i}\theta}v_{p}^{2}+2\mathrm{e}^{-\beta E_{p}}\mathrm{e}^{\mathrm{i}\theta}+\mathrm{e}^{-2\beta E_{p}}(\mathrm{e}^{2\mathrm{i}\theta}u_{p}^{2}+v_{p}^{2})}.
\end{align}
where $v^2$ and $u^2$ are the occupation and non-occupation numbers from  BCS solutions; and $\beta$ denotes $1/k_{\text{B}}T$ (the inverse of temperature);
$E$ denotes the quasi-particle energies.
At the limit of $\beta\rightarrow \infty$, these expressions can regain the expressions at zero temperature.
The procedure to derive these single-particle density matrices are provided in the appendix \ref{Appendix A}.

To calculate the free energy $F_{\text{p}} = E_{\text{p}} - TS_{\text{p}}$, the entropy must be determined. However, the exact calculation of entropy via $\mathop{\text{Tr}}(D_{\text{p}}\ln D_{\text{p}})$ is computationally intensive and impractical for heavy nuclei \cite{gambacurtaThermodynamicalPropertiesSmall2012}. Therefore, we adopt the method proposed in \cite{fantoParticlenumberProjectionFinitetemperature2017} to calculate the entropy, which is expressed as
\begin{equation}
    S_{\text{p}}=\ln(Z_{\text{p}})-\beta\frac{\partial\ln(Z_{\text{p}})}{\partial\beta}.
\end{equation}

This approximation is nearly exact when the pairing gap approaches zero as temperature increases but would approach to a negative number at low temperatures\cite{fantoParticlenumberProjectionFinitetemperature2017}. Consequently only the temperature above the critical temperature($T_{\text{c}}$) leads to accurate results.

To calculate the level density  with a particle number $N$, the saddle point approximation is employed, which is given by
\begin{align}
\rho_{N}(E^*) & =\left(2\mathrm{\pi}\left|\frac{\partial E_{N}}{\partial\beta}\right|\right)^{-1/2}Z_N\mathrm{e}^{\beta E_N},
\label{leveldensity}
\end{align}
where $E^*$ is the PNP energy difference between finite temperature and zero temperature. $E_N$ is given by
\begin{equation}
    E_N = -\frac{\partial \ln Z_N}{\partial \beta}\label{canonical energy}
\end{equation}
The canonical energy $E_N$ and entropy $S_N$ are calculated from the canonical partition function $Z_N$, which could be calculated by the DG method\cite{y.alhassidBenchmarkingMeanfieldApproximations2016} and the projection method\cite{fantoParticlenumberProjectionFinitetemperature2017}.
Note that $E_N$ can be seen as approximate excitation energies\cite{fantoParticlenumberProjectionFinitetemperature2017},
in contrast to exact PNP energies in Eq.(\ref{energy}).
 In the projection method, $Z_N$ is simply given by $Z_{\text{p}}\mathrm{e}^{\mathrm{\alpha}N}$, and $\alpha=-\beta \mu$ with the Fermi level $\mu$.
 In the DG method\cite{y.alhassidBenchmarkingMeanfieldApproximations2016}, the canonical partition function $Z_N$ is given by
\begin{equation}
    Z_{N}=\left(\sum_{N'}\exp\left(-\frac{1}{2}\frac{(N'-N)^{2}}{\langle\Delta N^{2}\rangle}\right)\right)^{-1}\mathrm{e}^{\alpha N}\Xi(\alpha,\beta).
\end{equation}
where $\Xi(\alpha,\beta)$ denotes the grand canonical partition function without projection.

Our PNP calculations are based on self-consistent Skyrme Hartree-Fock+BCS calculations.
Presently the SkM* force\cite{bartelBetterParametrisationSkyrmelike1982} is used in the particle-hole channel interaction, which is well-suited for nuclear calculations at large deformations.
In the pairing channel, we use the density-dependent $\delta$ interaction\cite{chasmanDensitydependentDeltaInteractions1976} in the mixed variant. The pairing strengths are set to  $V_\text{p} = -480\,\text{MeV}\,\text{fm}^{-3}$ and $V_\text{n} = -450\,\text{MeV}\,\text{fm}^{-3}$. The HF-BCS calculations at finite temperature are carried out in axial-symmetric coordinate spaces
 using the SkyAx code\cite{reinhardAxialHartreeFock2021}.
In PNP calculations, the divergence problem can appear when the denominator in gauge angle $\theta$  dependent densities becomes zero\cite{anguianoParticleNumberProjection2001}.
This issue also appears at finite temperature but happens very rarely.

\section{Results}

\begin{figure}[ht!]
    \centering
    \includegraphics[width=\linewidth]{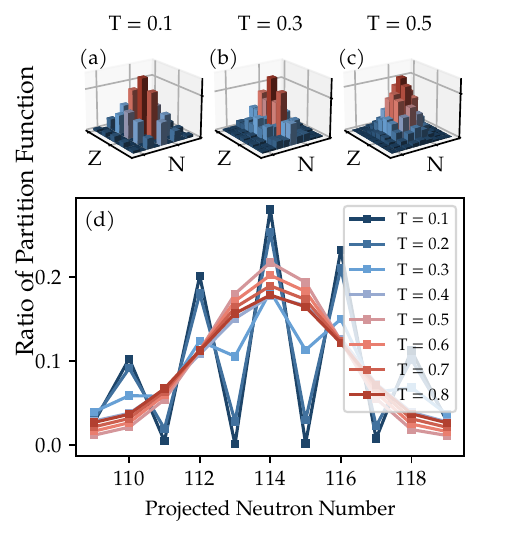}
    \caption{ The ratio of PNP partition function  $Z_{\text{p}}/\Xi$ for $\ce{^{292}Fl}$ at various temperatures.
    The ratio actually denotes the component proportion with specific particle number in the compound nucleus.
    (a, b, c) the ratios are shown for $T=0.1,0.3$ and $0.5\,\text{MeV}$, where the three-dimensional coordinates correspond to proton number, neutron number, and the ratio, respectively. (d) the ratios are shown in terms of different neutron numbers at various temperatures.
      All temperatures are given in MeV units.}
    \label{fig:figure1}
\end{figure}

First, we calculate the ratio $Z_{\text{p}}/\Xi$ of different particle numbers for $\ce{^{292}Fl}$, as shown in Fig.\ref{fig:figure1}.
The ratios in terms of proton and neutron numbers are shown in Fig.\ref{fig:figure1}(a,b,c).
The detailed ratios as a function of temperature and neutron numbers are shown in Fig.\ref{fig:figure1}(d).
The ratio denotes the component proportion with specific particle number in the compound nucleus, which is the analogy of $\langle\Psi|\hat{P}|\Psi\rangle$ at finite temperature\cite{sheikhSymmetryRestorationMeanfield2021}. At low temperatures, the superfluid system is dominated by even particle number components. As the temperature increases, the contribution of the odd particle number components begins to increase and the odd-even staggering structures disappear gradually. At $T = 0.4\,\text{MeV}$, the distribution of ratio has a Gaussian-like shape with a broad distribution. The distribution at $T=0.5\,\text{MeV}$ becomes narrow compared to $T=0.4\,\text{MeV}$. For temperatures above $T=0.5\,\text{MeV}$, the width of the Gaussian-like distribution increases, owing to the thermal effect.
Furthermore, we see that the ratios of components display non-monotonic changes due to the interplay of pairing effect and thermal effect.
For example, the ratio of $\ce{^{292}Fl}$ firstly decreases until $T=0.4\,\text{MeV}$, but rapidly increases at $T=0.5\,\text{MeV}$ followed by a decrease again.

\begin{figure}[ht!]
    \centering
    \includegraphics[width=\linewidth]{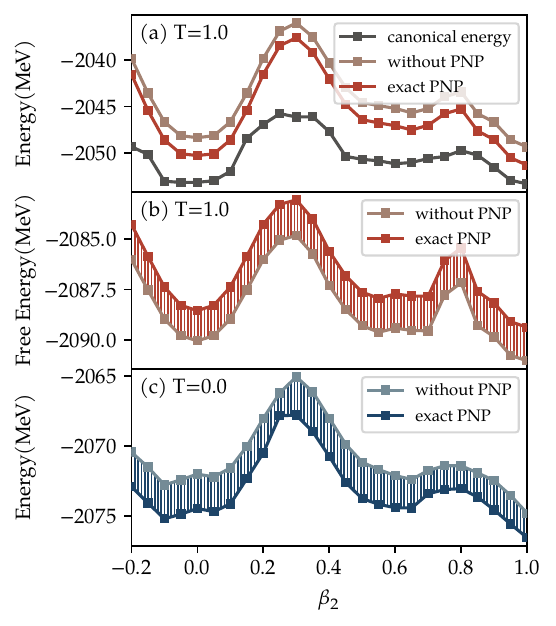}
    \caption{ (a) At temperature $T=1.0\,\text{MeV}$, the exact PNP energies, and energies without PNP, and the energies based on approximate canonical energies of $\ce{^{292}Fl}$
    as a function of quardpole deformation $\beta_2$.
    (b) the fission barriers by PNP energies and FT-BCS calculations at $T=1.0\,\text{MeV}$.
    (c) the fission barriers by PNP energies and Skyrme HF-BCS calculations at zero temperature. }
    \label{fig:figure2}
\end{figure}

Next the PNP energies are studied using different methods for comparison, which are the main results of this work.
Fig.\ref{fig:figure2}(a) shows the PNP energies of $\ce{^{292}Fl}$ as a function of quardpole deformation $\beta_2$ obtained by
exact PNP calculations using Eq.(\ref{energy}), the energies based on approximate canonical energies  using Eq.(\ref{canonical energy}) and
 energies from FT-BCS calculations without PNP, respectively.
It shows that the trend of exact PNP energies are similar to that without PNP at $T=1.0\,\text{MeV}$, but the exact PNP energies are systematically
lowered by about $2.0\,\text{MeV}$. This is understandable as PNP includes more correlations.
The approximate energies with excitation energies from  Eq.(\ref{canonical energy}) are lower than exact PNP energies by several MeVs.
In particular, the approximate energies at the barrier are significantly lower than exact PNP energies.
This demonstrates the essential role of exact PNP calculations of energies.
The fission barriers in terms of free energies are also shown in Fig.\ref{fig:figure2}(b).
The ground state shape at $T=1.0\,\text{MeV}$ becomes spherical due to the shape phase transition, while
it has a oblate shape at zero temperature.
We see the fission barrier at $T=1.0\,\text{MeV}$ with PNP and without PNP are similar, although
the free energies are systematically increased, due to the entropy by PNP are also decreased.
The barrier heights with PNP and without PNP are $5.47$ and $5.20\,\text{MeV}$, respectively.
The heights of fission barriers are sensitive in calculations of fission rates of compound nuclei\cite{qiaoModelingSurvivalProbabilities2022}.
For comparison, the fission barrier at zero temperature are also shown in Fig.\ref{fig:figure2}(c).
The barrier heights with PNP and without PNP are 7.38 and 7.69 MeV, respectively.
It can be seen that the PNP energies are systematically lower than that without PNP.
This is because the exact PNP at zero temperature has large influence when the pairing correlations are strong.
At high temperatures, we demonstrate that FT-BCS calculations of fission barriers are sufficiently close compared to
exact PNP calculations.

\begin{figure}[ht!]
    \centering
    \includegraphics[width=\linewidth]{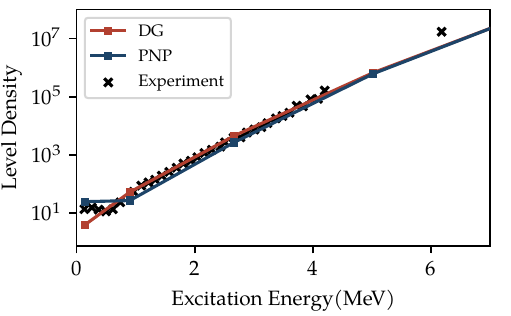}
    \caption{ Calculated level density of $\ce{^{238}U}$ with PNP partition function and the DG method.
    Note that the collective enhancement factors are included to compare with experimental data\cite{guttormsenConstanttemperatureLevelDensities2013}. }
    \label{fig:figure3}
\end{figure}

The energy dependent level densities from PNP are also studied in this work.
There have been extensive studies of level densities in \cite{goriely2008ImprovedMicroscopic,GORIELY2025139677, zhangLevelDensityOddA2023,jiangNuclearLevelDensity2024,wangNuclearLevelDensity2025,chenShellmodelbasedInvestigationLevel2023}.
Fig.\ref{fig:figure3} shows the calculated level densities of $\ce{^{238}U}$ at the low excitation region where experimental data is available.
The level densities are calculated using Eq.(\ref{leveldensity}) but with the PNP partition function and the DG method, respectively.
To compare with the experimental data, the  level densities are multiplied by collective enhancement factors $K_{\text{rot}}$ and $K_{\text{vib}}$ as defined in \cite{iljinovPhenomenologicalStatisticalAnalysis1992}.
We see that both results agree well with the experimental data in \cite{guttormsenConstanttemperatureLevelDensities2013}.
However, the DG method  shows obvious deviations from the experimental data before $E_x<1\,\text{MeV}$, corresponding to $T=0.3\,\text{MeV}$.
This discrepancy arises because the DG method assume a Gaussian-shape ratio between the grand canonical partition function $\Xi(\alpha)$ and the canonical partition function $Z_N$, which are discussed in the appendix \ref{Appendix B}. From the perspective of the projection method, as illustrated in \ref{fig:figure1}, the pairing interaction induces odd-even staggering behavior in the ratio distribution at low temperatures, thereby invalidating the DG assumption at low temperatures.

\begin{figure}[ht!]
    \centering
    \includegraphics[width=\linewidth]{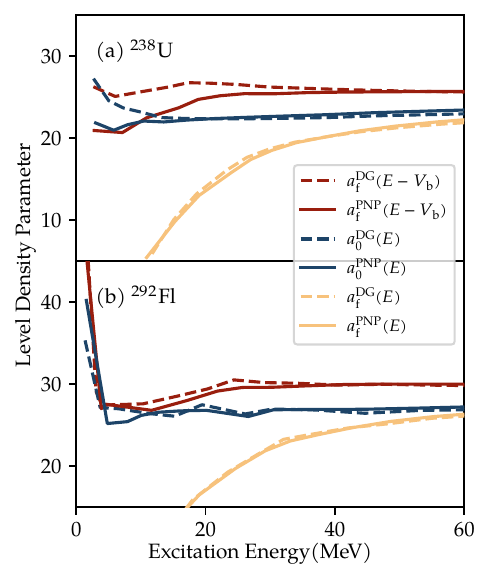}
    \caption{ The extracted level density parameters for $\ce{^{238}U}$ and  $\ce{^{292}Fl}$
    based on the level densities from PNP partition function and the DG method.
    The level densities at the ground state deformation and the barrier
    are given as $a_{\text{0}}(E)$ and $a_{\text{f}}(E-V_\text{b})$, respectively.
    $a_{\text{f}}(E)$ are also shown to compare with $a_{\text{0}}(E)$.}
    \label{fig:figure4}
\end{figure}

The level density parameter $a$ can be extracted from our results of level densities using the back-shifted Fermi gas model\cite{dilgLevelDensityParameters1973}, as shown in Fig.\ref{fig:figure4}.
The level density parameter is an empirical parameter and has been widely used in statistical calculations of survival probabilities of compound nuclei\cite{qiangSurvivalProbabilitiesCompound2024,qiaoModelingSurvivalProbabilities2022}.
Here the collective enhancement factors are not considered to extract $a$ to reduce the model dependence.
Generally level density parameters from DG are slightly larger than that from PNP and they are close at high excitation energies.
It can be seen that the level density parameter at the barrier, $a_{\text{f}}$,  is larger than that at the ground state $a_{\text{0}}$ at high excitation energies.
Note that the level density parameter at the saddle point is displayed as  $a_{\text{f}}(E-V_{\text{b}})$, where $V_{\text{b}}$ is the barrier height at zero temperature.
At low excitation energies, the obtained  $a_{\text{f}}$ is close to $a_{\text{0}}$, which increases with increasing excitation energies and becomes stable above $30\,\text{MeV}$.
For $\ce{^{238}U}$ and $\ce{^{292}Fl}$, the rations $a_{\text{f}}/a_{\text{0}}$ are $1.0718$ and $1.0625$ at high excitation energies, respectively.
Note that $a_{\text{f}}/a_{\text{0}}$ is usually adopted around $1.1$ as a key adjustable parameter in the literature\cite{fengProductionCrossSections2006,qiangSurvivalProbabilitiesCompound2024}.
The level density parameter $a_{\text{0}}$ is around $A/9.97$, $A/10.38$ for $\ce{^{238}U}$ and $\ce{^{292}Fl}$, respectively.
Usually the level density parameter $a_{\text{0}}$ is taken around $A/10 \sim A/12$ in statistical models\cite{fengProductionCrossSections2006,zubovSurvivalProbabilitySuperheavy2002,xiaSystematicStudySurvival2011,zhangPredictionsSynthesizingElements2024,ignatyukFissionPreactinideNuclei1975,dengExaminationPromisingReactions2023}.
Our microscopic results of $a_{\text{0}}$ and $a_{\text{f}}/a_{\text{0}}$ can provide useful guidance for statistical models, which is a longstanding issue.
The  $a_{\text{f}}(E)$ without subtracting the barrier height are also shown to compare with $a_{\text{0}}(E)$.
In this case, $a_{\text{f}}(E)$ is not physical below the barrier. We see that $a_{\text{f}}(E)$ becomes close to $a_{\text{0}}(E)$ around $60\,\text{MeV}$, indicating
the meltdown  of shell effects.

\section{Summary}

To summarize, we have completed a comprehensive particle number projection formalism for evaluating physics operators at finite temperature.
In particular, the exact PNP energies of compound nuclei can now be calculated, based on the self-consistent Skyrme DFT with pairing correlations.
The temperature dependent partition function shows that the odd-even staggering structures in terms of different particle numbers at low temperatures,
 which become Gaussian-shape structures at high temperatures.
 The PNP energies and free energies as a function of deformations are calculated,
 which shows that the fission barriers are close to results without PNP at high temperatures.
 The approximate canonical energies are significantly different to exact PNP energies especially at the barrier.
 The level densities are also studied and agree with experimental data when the collective enhancement factors are included.
 Finally the level density parameters and the ratio $a_{\text{f}}/a_{\text{0}}$ are extracted from the level densities of  $\ce{^{238}U}$ and $\ce{^{292}Fl}$,
 which are useful for constrain statical models to calculate the survival probabilities of compound superheavvy nuclei.
 Note that the exact calculations of PNP entropy is still computationally too costly.
 The present method applied to PNP energies at finite temperature could also be useful to calculate other observables.

 \acknowledgments
 This work was supported by  the
 National Key R$\&$D Program of China (Grant No.2023YFE0101500,2023YFA1606403),
  the National Natural Science Foundation of China under Grants No.12475118, 12335007.

\appendix

\section{Calculations of $\rho, \kappa, \lambda$} \label{Appendix A}

First, the creation and annihilation operators can be extracted from the exponent and expressed explicitly outside the exponential form by

\begin{equation}
        \begin{aligned}
            &\mathrm{e}^{\mathrm{i}\theta\hat{N}}=\prod_{p}\mathrm{e}^{\mathrm{i}\theta a_{p}^{\dagger}a_{p}}=\prod_{p}[1+(\mathrm{e}^{\mathrm{i}\theta}-1)a_{p}^{\dagger}a_{p}]\\
            =&\prod_{p>0}[1+(\mathrm{e}^{\mathrm{i}\theta}-1)(a_{p}^{\dagger}a_{p}+a_{\bar{p}}^{\dagger}a_{\bar{p}})+(\mathrm{e}^{\mathrm{i}\theta}-1)^{2}a_{p}^{\dagger}a_{p}a_{\bar{p}}^{\dagger}a_{\bar{p}}].
        \end{aligned}
\end{equation}

Then the single-particle creation and annihilation operators can be replaced by quasi-particle creation and annihilation operators using

\begin{equation}
    \begin{aligned}
        \mathrm{e}^{\mathrm{i}\theta\hat{N}} &=\prod_{p>0}[(\mathrm{e}^{\mathrm{i}\theta}-1)^{2}\alpha_{p}^{\dagger}\alpha_{p}\alpha_{\bar{p}}^{\dagger}\alpha_{\bar{p}}+b_{p}(\alpha_{p}^{\dagger}\alpha_{p}+\alpha_{\bar{p}}^{\dagger}\alpha_{\bar{p}})\\
        &+c_{p}(\alpha_{p}^{\dagger}\alpha_{\bar{p}}^{\dagger}+\alpha_{\bar{p}}\alpha_{p})+d_{p}],
    \end{aligned}
\end{equation}

where

\begin{equation}
    \begin{aligned}
        b_{p} & =\mathrm{e}^{\mathrm{i}\theta}-1+(1-\mathrm{e}^{2\mathrm{i}\theta})v_{p}^{2},\\
        c_{p} & =(\mathrm{e}^{2\mathrm{i}\theta}-1)u_{p}v_{p},\\
        d_{p} & =\mathrm{e}^{2\mathrm{i}\theta}v_{p}^{2}+u_{p}^{2}.
    \end{aligned}
\end{equation}

Now we can write $\mathrm{e}^{\mathrm{i}\theta\hat{N}}$ in quasi-particle number representation and thus calculate $\text{Tr}(D\mathrm{e}^{\mathrm{i}\theta\hat{N}})$ by classical statistic mechanics trick. For convenience, we let $D = \mathrm{e}^{-\beta\hat{K}}$, where $\hat{K}=\hat{H}_{\text{BCS}}-\mu\hat{N}=\sum_{p}E_{p}\alpha^{\dagger}_{p}\alpha_{p}$:

\begin{equation}
    \begin{aligned}
        &\mathop{\text{Tr}}(D\mathrm{e}^{\mathrm{i}\theta\hat{N}}) =\sum_{\{n_{p}\}}\langle\{n_{p}\}|\mathrm{e}^{-\beta K}\mathrm{e}^{\mathrm{i}\theta\hat{N}}|\{n_{p}\}\rangle\\
        =&\sum_{\{n_{p}\}}\mathrm{e}^{-\beta\sum_{p}n_{p} E_{p}}\langle\{n_{p}\}|\mathrm{e}^{\mathrm{i}\theta\hat{N}}|\{n_{p}\}\rangle\\
        =&\sum_{\{n_{p}\}}\prod_{p>0}\mathrm{e}^{-\beta(n_{p}+n_{\bar{p}}) E_{p}}[(\mathrm{e}^{\mathrm{i}\theta}-1)^{2}n_{p}n_{\bar{p}}+b_{p}(n_{p}+n_{\bar{p}})+d_{p}]\\
        =&\prod_{p>0}[u_{p}^{2}+\mathrm{e}^{2\mathrm{i}\theta}v_{p}^{2}+2\mathrm{e}^{-\beta E_{p}}\mathrm{e}^{\mathrm{i}\theta}+\mathrm{e}^{-2\beta E_{p}}(\mathrm{e}^{2\mathrm{i}\theta}u_{p}^{2}+v_{p}^{2})].
    \end{aligned}
\end{equation}

Similarly, we can calculate $\mathop{\text{Tr}}(D\alpha_{p}^{\dagger}\alpha_{q}\mathrm{e}^{\mathrm{i}\theta\hat{N}})$, $\mathop{\text{Tr}}(D\alpha_{p}^{\dagger}\alpha_{q}^{\dagger}\mathrm{e}^{\mathrm{i}\theta\hat{N}})$ and $\mathop{\text{Tr}}(D\alpha_{p}\alpha_{q}\mathrm{e}^{\mathrm{i}\theta\hat{N}})$, thus we can get

\begin{equation}
    \begin{aligned}
        & \langle\alpha_{p}^{\dagger}\alpha_{q}\rangle_{\theta}\equiv\frac{\mathop{\text{Tr}}(D\alpha_{p}^{\dagger}\alpha_{q}\mathrm{e}^{\mathrm{i}\theta\hat{N}})}{\mathop{\text{Tr}}(D\mathrm{e}^{\mathrm{i}\theta\hat{N}})}\\
        = & \delta_{pq}\frac{\mathrm{e}^{-\beta E_{p}}\mathrm{e}^{\mathrm{i}\theta}+\mathrm{e}^{-2\beta E_{p}}(\mathrm{e}^{2\mathrm{i}\theta}u_{p}^{2}+v_{p}^{2})}{u_{p}^{2}+\mathrm{e}^{2\mathrm{i}\theta}v_{p}^{2}+2\mathrm{e}^{-\beta E_{p}}\mathrm{e}^{\mathrm{i}\theta}+\mathrm{e}^{-2\beta E_{p}}(\mathrm{e}^{2\mathrm{i}\theta}u_{p}^{2}+v_{p}^{2})},\\
        &\langle\alpha_{p}^{\dagger}\alpha_{\bar{q}}^{\dagger}\rangle_{\theta}\equiv\frac{\mathop{\text{Tr}}(D\alpha_{p}^{\dagger}\alpha_{\bar{q}}^{\dagger}\mathrm{e}^{\mathrm{i}\theta\hat{N}})}{\mathop{\text{Tr}}(D\mathrm{e}^{\mathrm{i}\theta\hat{N}})}\\
        = & \delta_{pq}\frac{\mathrm{e}^{-2\beta E_{p}}(\mathrm{e}^{2\mathrm{i}\theta}-1)u_{p}v_{p}}{u_{p}^{2}+\mathrm{e}^{2\mathrm{i}\theta}v_{p}^{2}+2\mathrm{e}^{-\beta E_{p}}\mathrm{e}^{\mathrm{i}\theta}+\mathrm{e}^{-2\beta E_{p}}(\mathrm{e}^{2\mathrm{i}\theta}u_{p}^{2}+v_{p}^{2})},\\
         &\langle\alpha_{\bar{p}}\alpha_{q}\rangle_{\theta}\equiv\frac{\mathop{\text{Tr}}(D\alpha_{\bar{p}}\alpha_{q}\mathrm{e}^{\mathrm{i}\theta\hat{N}})}{\mathop{\text{Tr}}(D\mathrm{e}^{\mathrm{i}\theta\hat{N}})}\\
        = & \delta_{pq}\frac{(\mathrm{e}^{2\mathrm{i}\theta}-1)u_{p}v_{p}}{u_{p}^{2}+\mathrm{e}^{2\mathrm{i}\theta}v_{p}^{2}+2\mathrm{e}^{-\beta E_{p}}\mathrm{e}^{\mathrm{i}\theta}+\mathrm{e}^{-2\beta E_{p}}(\mathrm{e}^{2\mathrm{i}\theta}u_{p}^{2}+v_{p}^{2})}.
    \end{aligned}
\end{equation}

Using the BCS transformation, we can get the final results:

\begin{equation}
    \begin{aligned}
        \tilde{\rho}_{qp} & =\delta_{pq}\frac{\mathrm{e}^{-2\beta E_{p}}\mathrm{e}^{2\mathrm{i}\theta}u_{p}^{2}+\mathrm{e}^{-\beta E_{p}}\mathrm{e}^{\mathrm{i}\theta}+\mathrm{e}^{2\mathrm{i}\theta}v_{p}^{2}}{u_{p}^{2}+\mathrm{e}^{2\mathrm{i}\theta}v_{p}^{2}+2\mathrm{e}^{-\beta E_{p}}\mathrm{e}^{\mathrm{i}\theta}+\mathrm{e}^{-2\beta E_{p}}(\mathrm{e}^{2\mathrm{i}\theta}u_{p}^{2}+v_{p}^{2})},\\
        \tilde{\kappa}_{qp} & =\delta_{p\bar{q}}\frac{\mathrm{e}^{-2\beta E_{p}}\mathrm{e}^{2\mathrm{i}\theta}u_{p}v_{p}-\mathrm{e}^{2\mathrm{i}\theta}u_{p}v_{p}}{u_{p}^{2}+\mathrm{e}^{2\mathrm{i}\theta}v_{p}^{2}+2\mathrm{e}^{-\beta E_{p}}\mathrm{e}^{\mathrm{i}\theta}+\mathrm{e}^{-2\beta E_{p}}(\mathrm{e}^{2\mathrm{i}\theta}u_{p}^{2}+v_{p}^{2})},\\
        \tilde{\lambda}_{pq} & =\delta_{p\bar{q}}\frac{-\mathrm{e}^{-2\beta E_{p}}u_{p}v_{p}+u_{p}v_{p}}{u_{p}^{2}+\mathrm{e}^{2\mathrm{i}\theta}v_{p}^{2}+2\mathrm{e}^{-\beta E_{p}}\mathrm{e}^{\mathrm{i}\theta}+\mathrm{e}^{-2\beta E_{p}}(\mathrm{e}^{2\mathrm{i}\theta}u_{p}^{2}+v_{p}^{2})}.
    \end{aligned}
\end{equation}

\section{Derivation of Discrete Gaussian Approximation}\label{Appendix B}

We begin by making the assumption that the density of states $\rho(N, E)$ is non-zero only in the vicinity of $N = N_{0}$. Consequently, $Z_{N}$ also takes non-negligible values only near $N = N_{0}$. Based on this assumption, we approximate the distribution using a Gaussian form:

\begin{equation}
    \begin{aligned}
         &Z_{N} = \mathrm{e}^{\ln Z_{N}} \\
        =& Z_{N_{0}} \exp\left(  \frac{\partial \ln Z_{N}}{\partial N}  (N - N_{0}) + \frac{1}{2} \frac{\partial^{2} \ln Z_{N}}{\partial N^{2}} (N - N_{0})^{2} \right).
    \end{aligned}
\end{equation}

Thus, the grand canonical partition function can be expressed as:

\begin{equation}
    \begin{aligned}
        &\Xi(\alpha', \beta) = \sum_{N} \mathrm{e}^{-\alpha' N} Z_{N}(\beta) \\
        =& \mathrm{e}^{-\alpha' N_{0}} Z_{N_{0}}(\beta) \sum_{N} \exp\left( \frac{1}{2} \frac{\partial^{2} \ln Z_{N}}{\partial N^{2}} (N - N_{0})^{2} \right).
    \end{aligned}
\end{equation}

To eliminate the linear term in $N$, we set:

\begin{align}
    \left(\frac{\partial \ln Z_{N}}{\partial N} \right)_{N_{0}} = \alpha',
\end{align}

which constitutes the so-called discrete Gaussian approximation. However, to derive the canonical partition function from the grand canonical partition function, we further require:

\begin{equation}
    \begin{aligned}
        &\left(\frac{\partial^{2} \ln Z_{N}}{\partial N^{2}} \right)_{N_{0}} = \left(\frac{\partial \alpha'}{\partial N} \right)_{N_{0}} \simeq \left( \frac{\partial \alpha}{\partial N} \right)_{N_{0}} \\
        =& \left( \frac{\partial N}{\partial \alpha} \right)^{-1}_{\alpha_{0}} = -\left( \frac{\partial^{2} \ln \Xi}{\partial \alpha^{2}} \right)^{-1}_{\alpha_{0}}.
    \end{aligned}
\end{equation}

Here, the second step involves replacing $\alpha'$ with the chemical potential $\alpha$. It is important to note that the notation after the approximation generally implies that $\alpha_0$ corresponds to the chemical potential for particle number $N_0$, whereas the expression before the approximation is merely a symbolic substitution. Thus, this substitution is, in fact, an approximation. It can be demonstrated that this approximation is well-justified within our framework. The value $\alpha_{0}$ satisfies:

\begin{align}
    \left( \frac{\partial \ln \Xi}{\partial \alpha} \right)_{\alpha_{0}} = N_{0}.
\end{align}

We now prove that the above relation also holds approximately for $\alpha'$. By computing

\begin{equation}
    \begin{aligned}
        -\left( \frac{\partial \ln \Xi}{\partial \alpha} \right)_{\alpha'} &= \sum_{N} N \mathrm{e}^{-\alpha' N} Z_{N_{0}} \mathrm{e}^{\alpha' (N - N_{0}) + \frac{\beta'}{2} (N - N_{0})^{2}} \\
        &= \frac{ \sum_{N} N \mathrm{e}^{\frac{\beta'}{2} (N - N_{0})^{2}} }{ \sum_{N} \mathrm{e}^{\frac{\beta'}{2} (N - N_{0})^{2}} } \\
        &= \frac{ \sum_{N} (N - N_{0}) \mathrm{e}^{\frac{\beta'}{2} (N - N_{0})^{2}} }{ \sum_{N} \mathrm{e}^{\frac{\beta'}{2} (N - N_{0})^{2}} } + N_{0}
    \end{aligned}
\end{equation}

and assuming that $N_{0}$ is sufficiently large compared to the range where the Gaussian distribution contributes significantly, we can extend the summation over $N$ to negative infinity. Consequently, the first term in the above expression vanishes.
Finally the expression of $\Xi$ can be written as:

\begin{align}
    \Xi(\alpha,\beta) = \mathrm{e}^{-\alpha N_0}Z_{N_0}(\beta)\sum_N \exp\left(-\frac{1}{2}\frac{(N-N_0)^2}{\partial^2\ln\Xi/\partial\alpha^2}\right)
\end{align}

\newpage

\bibliography{main}

@article{benderSelfconsistentMeanfieldModels2003,
  author =        {Bender, Michael and Heenen, Paul-Henri and
                   Reinhard, Paul-Gerhard},
  journal =       {Rev. Mod. Phys.},
  month =         jan,
  number =        {1},
  pages =         {121--180},
  title =         {Self-Consistent Mean-Field Models for Nuclear
                   Structure},
  volume =        {75},
  year =          {2003},
  doi =           {10.1103/RevModPhys.75.121},
  issn =          {0034-6861, 1539-0756},
}

@article{qiangSurvivalProbabilitiesCompound2024,
  author =        {Qiang, Yu and Deng, Xiang Quan and Shi, Yue and
                   Qiao, Chun Yuan and Pei, Jun Chen},
  journal =       {Phys. Lett. B},
  month =         nov,
  pages =         {139057},
  title =         {Survival Probabilities of Compound Superheavy Nuclei
                   towards Element 119},
  volume =        {858},
  year =          {2024},
  doi =           {10.1016/j.physletb.2024.139057},
  issn =          {0370-2693},
}

@article{scampsImpactPearshapedFission2018,
  author =        {Scamps, Guillaume and Simenel, C{\'e}dric},
  journal =       {Nature},
  month =         dec,
  number =        {7736},
  pages =         {382--385},
  publisher =     {{Nature Publishing Group}},
  title =         {Impact of Pear-Shaped Fission Fragments on
                   Mass-Asymmetric Fission in Actinides},
  volume =        {564},
  year =          {2018},
  doi =           {10.1038/s41586-018-0780-0},
  issn =          {1476-4687},
}

@article{bulgacTimeDependentDensityFunctional2025,
  author =        {Bulgac, Aurel and Abdurrahman, Ibrahim and
                   Kafker, Matthew and Stetcu, Ionel},
  journal =       {Phys. Rev. Lett.},
  month =         aug,
  number =        {6},
  pages =         {062501},
  publisher =     {{American Physical Society}},
  title =         {Time-{{Dependent Density Functional Theory
                   Description}} of \$\^\{238\}\textbackslash
                   mathrm\{\vphantom\}{{U}}\vphantom\{\}(\textbackslash
                   mathrm\{n\},\textbackslash mathrm\{f\})\$,
                   \$\^\{240,242\}\textbackslash
                   mathrm\{\vphantom\}{{Pu}}\vphantom\{\}(\textbackslash
                   mathrm\{n\},\textbackslash mathrm\{f\})\$, and
                   \$\^\{237\}\textbackslash
                   mathrm\{\vphantom\}{{Np}}\vphantom\{\}(\textbackslash
                   mathrm\{n\},\textbackslash mathrm\{f\})\$
                   {{Reactions}}},
  volume =        {135},
  year =          {2025},
  doi =           {10.1103/2k8k-vpng},
}

@article{qiangQuantumEntanglementNuclear2025,
  author =        {Qiang, Yu and Pei, Jun Chen and Godbey, Kyle},
  journal =       {Phys. Lett. B},
  month =         feb,
  pages =         {139248},
  title =         {Quantum Entanglement in Nuclear Fission},
  volume =        {861},
  year =          {2025},
  doi =           {10.1016/j.physletb.2025.139248},
  issn =          {03702693},
}

@article{stoitsovLargescaleMassTable2009,
  author =        {Stoitsov, M. and Nazarewicz, W. and Schunck, N.},
  journal =       {Int. J. Mod. Phys. E},
  month =         apr,
  number =        {04},
  pages =         {816--822},
  publisher =     {{World Scientific Publishing Co.}},
  title =         {Large-Scale Mass Table Calculations},
  volume =        {18},
  year =          {2009},
  doi =           {10.1142/S0218301309012914},
  issn =          {0218-3013},
}

@article{guanHighQualityMicroscopic2024,
  author =        {Guan, Da Wei and Pei, Jun Chen},
  journal =       {Phys. Lett. B},
  month =         apr,
  pages =         {138578},
  title =         {High Quality Microscopic Nuclear Masses of Superheavy
                   Nuclei},
  volume =        {851},
  year =          {2024},
  doi =           {10.1016/j.physletb.2024.138578},
  issn =          {0370-2693},
}

@article{zhouNeutronHaloDeformed2010,
  author =        {Zhou, Shan Gui and Meng, Jie and Ring, P. and
                   Zhao, En Guang},
  journal =       {Phys. Rev. C},
  month =         jul,
  number =        {1},
  pages =         {011301},
  publisher =     {{American Physical Society}},
  title =         {Neutron Halo in Deformed Nuclei},
  volume =        {82},
  year =          {2010},
  doi =           {10.1103/PhysRevC.82.011301},
}

@article{pei2005density,
  author =        {Pei, Jun Chen and Xu, Fu Rong and Stevenson, P. D.},
  journal =       {Phys. Rev. C},
  number =        {3},
  pages =         {034302},
  publisher =     {APS},
  title =         {Density distributions of superheavy nuclei},
  volume =        {71},
  year =          {2005},
  doi =           {10.1103/PhysRevC.71.034302},
}

@article{wardaFissionHalflivesSuperheavy2012,
  author =        {Warda, M. and Egido, J. L.},
  journal =       {Phys. Rev. C},
  month =         jul,
  number =        {1},
  pages =         {014322},
  publisher =     {{American Physical Society}},
  title =         {Fission Half-Lives of Superheavy Nuclei in a
                   Microscopic Approach},
  volume =        {86},
  year =          {2012},
  doi =           {10.1103/PhysRevC.86.014322},
}

@article{marevicFissionPu2402020,
  author =        {Marevi{\'c}, P. and Schunck, N.},
  journal =       {Phys. Rev. Lett.},
  month =         sep,
  number =        {10},
  pages =         {102504},
  title =         {Fission of {{Pu}} 240 with {{Symmetry-Restored
                   Density Functional Theory}}},
  volume =        {125},
  year =          {2020},
  doi =           {10.1103/PhysRevLett.125.102504},
  issn =          {0031-9007, 1079-7114},
}

@article{nSchunck2014PotentialEnergySurface,
  author =        {Schunck, N. and Duke, D. and Carr, H. and Knoll, A.},
  journal =       {Phys. Rev. C},
  month =         {Nov},
  pages =         {054305},
  publisher =     {American Physical Society},
  title =         {Description of induced nuclear fission with Skyrme
                   energy functionals: Static potential energy surfaces
                   and fission fragment properties},
  volume =        {90},
  year =          {2014},
  doi =           {10.1103/PhysRevC.90.054305},
  url =           {https://link.aps.org/doi/10.1103/PhysRevC.90.054305},
}

@article{chi2023role,
  author =        {Chi, Ji Huai and Qiang, Yu and Gao, Chun Yuan and
                   Pei, Jun Chen},
  journal =       {Nucl. Phys. A},
  pages =         {122626},
  publisher =     {Elsevier},
  title =         {Role of hexadecapole deformation in fission potential
                   energy surfaces of 240Pu},
  volume =        {1032},
  year =          {2023},
  doi =           {10.1016/j.nuclphysa.2023.122626},
}

@article{nazarewiczTheoreticalDescriptionSuperheavy2002,
  author =        {Nazarewicz, W. and Bender, M. and {\'C}wiok, S. and
                   Heenen, P. H. and Kruppa, A. T. and Reinhard, P. -G. and
                   Vertse, T.},
  journal =       {Nucl. Phys. A},
  month =         apr,
  number =        {1},
  pages =         {165--171},
  title =         {Theoretical Description of Superheavy Nuclei},
  volume =        {701},
  year =          {2002},
  doi =           {10.1016/S0375-9474(01)01567-6},
  issn =          {0375-9474},
}

@article{alanl.goodmanFinitetemperatureHFBTheory1981,
  author =        {{A. L. Goodman}},
  journal =       {Nucl. Phys.},
  month =         jan,
  number =        {1},
  pages =         {30--44},
  title =         {Finite-Temperature {{HFB}} Theory},
  volume =        {352},
  year =          {1981},
  doi =           {10.1016/0375-9474(81)90557-1},
}

@article{peiFissionBarriersCompound2009,
  author =        {Pei, J. C. and Nazarewicz, W. and Sheikh, J. A. and
                   Kerman, A. K.},
  journal =       {Phys. Rev. Lett.},
  month =         may,
  number =        {19},
  pages =         {192501},
  title =         {Fission {{Barriers}} of {{Compound Superheavy
                   Nuclei}}},
  volume =        {102},
  year =          {2009},
  doi =           {10.1103/PhysRevLett.102.192501},
  issn =          {0031-9007, 1079-7114},
}

@article{sheikhSymmetryRestorationMeanfield2021,
  author =        {Sheikh, J. A. and Dobaczewski, J. and Ring, P. and
                   Robledo, L. M. and Yannouleas, C.},
  journal =       {J. Phys. G},
  month =         dec,
  number =        {12},
  pages =         {123001},
  title =         {Symmetry Restoration in Mean-Field Approaches},
  volume =        {48},
  year =          {2021},
  doi =           {10.1088/1361-6471/ac288a},
  issn =          {0954-3899, 1361-6471},
}

@book{ringNuclearManyBody2004,
  address =       {{Berlin Heidelberg}},
  author =        {Ring, Peter and Schuck, Peter},
  edition =       {1. ed., 3. print., study ed},
  publisher =     {{Springer}},
  title =         {The Nuclear Many Body Problem},
  year =          {2004},
  isbn =          {978-3-540-21206-5},
}

@article{m.v.stoitsovVariationParticlenumberProjection2007,
  author =        {{M. V. Stoitsov} and {J. Dobaczewski} and
                   {R. Kirchner} and {W. Nazarewicz} and {J. Terasaki}},
  journal =       {Phys. Rev. C},
  month =         jul,
  number =        {1},
  pages =         {014308},
  title =         {Variation after Particle-Number Projection for the
                   {{Hartree-Fock-Bogoliubov}} Method with the
                   {{Skyrme}} Energy Density Functional},
  volume =        {76},
  year =          {2007},
  doi =           {10.1103/physrevc.76.014308},
}

@article{bertschSymmetryRestorationHartreeFockBogoliubov2012,
  author =        {Bertsch, G. F. and Robledo, L. M.},
  journal =       {Phys. Rev. Lett.},
  month =         jan,
  number =        {4},
  pages =         {042505},
  title =         {Symmetry {{Restoration}} in {{Hartree-Fock-Bogoliubov
                   Based Theories}}},
  volume =        {108},
  year =          {2012},
  doi =           {10.1103/PhysRevLett.108.042505},
  issn =          {0031-9007, 1079-7114},
}

@article{bulgacProjectionGoodQuantum2019,
  author =        {Bulgac, Aurel},
  journal =       {Phys. Rev. C},
  month =         sep,
  number =        {3},
  pages =         {034612},
  title =         {Projection of Good Quantum Numbers for Reaction
                   Fragments},
  volume =        {100},
  year =          {2019},
  doi =           {10.1103/PhysRevC.100.034612},
  issn =          {2469-9985, 2469-9993},
}

@article{verriereMicroscopicCalculationFission2021,
  author =        {Verriere, Marc and Schunck, Nicolas and
                   Regnier, David},
  journal =       {Phys. Rev. C},
  month =         may,
  number =        {5},
  pages =         {054602},
  title =         {Microscopic Calculation of Fission Product Yields
                   with Particle-Number Projection},
  volume =        {103},
  year =          {2021},
  doi =           {10.1103/PhysRevC.103.054602},
  issn =          {2469-9985, 2469-9993},
}

@article{ballyProjectionParticleNumber2021,
  author =        {Bally, Benjamin and Bender, Michael},
  journal =       {Phys. Rev. C},
  month =         feb,
  number =        {2},
  pages =         {024315},
  title =         {Projection on Particle Number and Angular Momentum:
                   {{Example}} of Triaxial {{Bogoliubov}} Quasiparticle
                   States},
  volume =        {103},
  year =          {2021},
  doi =           {10.1103/PhysRevC.103.024315},
  issn =          {2469-9985, 2469-9993},
}

@article{rossignoliProjectionFiniteTemperature1994,
  author =        {Rossignoli, R. and Ring, P.},
  journal =       {Ann. Phys.},
  month =         nov,
  number =        {2},
  pages =         {350--389},
  title =         {Projection at {{Finite Temperature}}},
  volume =        {235},
  year =          {1994},
  doi =           {10.1006/aphy.1994.1101},
  issn =          {0003-4916},
}

@article{jiangQuantumComputingPairing2023,
  author =        {Jiang, Chong Ji and Pei, Jun Chen},
  journal =       {Phys. Rev. C},
  number =        {4},
  pages =         {044308},
  title =         {Quantum Computing of the Pairing {{Hamiltonian}} at
                   Finite Temperature},
  volume =        {107},
  year =          {2023},
  doi =           {10.1103/PhysRevC.107.044308},
}

@article{y.alhassidBenchmarkingMeanfieldApproximations2016,
  author =        {{Y. Alhassid} and {G. F. Bertsch} and
                   {C. N. Gilbreth} and {H. Nakada}},
  journal =       {Phys. Rev. C},
  month =         apr,
  number =        {4},
  pages =         {044320},
  title =         {Benchmarking Mean-Field Approximations to Level
                   Densities},
  volume =        {93},
  year =          {2016},
  doi =           {10.1103/physrevc.93.044320},
}

@article{nogamiImprovedSuperconductivityApproximation1964,
  author =        {Nogami, Yukihisa},
  journal =       {Phys. Rev.},
  month =         apr,
  number =        {2B},
  pages =         {B313-B321},
  publisher =     {{American Physical Society}},
  title =         {Improved {{Superconductivity Approximation}} for the
                   {{Pairing Interaction}} in {{Nuclei}}},
  volume =        {134},
  year =          {1964},
  doi =           {10.1103/PhysRev.134.B313},
}

@article{wang2014Lipkin,
  author =        {Wang, X. B. and Dobaczewski, J. and Kortelainen, M. and
                   Yu, L. F. and Stoitsov, M. V.},
  journal =       {Phys. Rev. C},
  month =         {Jul},
  pages =         {014312},
  publisher =     {American Physical Society},
  title =         {Lipkin method of particle-number restoration to
                   higher orders},
  volume =        {90},
  year =          {2014},
  doi =           {10.1103/PhysRevC.90.014312},
  url =           {https://link.aps.org/doi/10.1103/PhysRevC.90.014312},
}

@article{sheikhSymmetryprojectedHartreeFock2000,
  author =        {Sheikh, Javid A. and Ring, Peter},
  journal =       {Nucl. Phys. A},
  month =         feb,
  number =        {1},
  pages =         {71--91},
  title =         {Symmetry-Projected
                       {{Hartree}}\textendash{{Fock}}\textendash{{Bogoliubov}}
  Equations},
  volume =        {665},
  year =          {2000},
  doi =           {10.1016/S0375-9474(99)00424-8},
  issn =          {0375-9474},
}

@article{anguianoParticleNumberProjection2001,
  author =        {Anguiano, M. and Egido, J. L. and Robledo, L. M.},
  journal =       {Nucl. Phys. A},
  month =         dec,
  number =        {3},
  pages =         {467--493},
  title =         {Particle Number Projection with Effective Forces},
  volume =        {696},
  year =          {2001},
  doi =           {10.1016/S0375-9474(01)01219-2},
  issn =          {0375-9474},
}

@article{rossignoliFiniteTemperatureProjected1992,
  author =        {Rossignoli, R. and Ring, P. and Dinh Dang, N.},
  journal =       {Phys. Lett. B},
  month =         dec,
  number =        {1-2},
  pages =         {9--13},
  title =         {Finite Temperature Projected Calculations in the
                   Static Path Approximation},
  volume =        {297},
  year =          {1992},
  doi =           {10.1016/0370-2693(92)91060-M},
  issn =          {03702693},
}

@article{esebbagNumberProjectedStatistics1993,
  author =        {Esebbag, C. and Egido, J. L.},
  journal =       {Nucl. Phys. A},
  month =         feb,
  number =        {2},
  pages =         {205--231},
  title =         {Number Projected Statistics and the Pairing
                   Correlations at High Excitation Energies},
  volume =        {552},
  year =          {1993},
  doi =           {10.1016/0375-9474(93)90464-9},
  issn =          {03759474},
}

@article{rossignoliProjectedStatisticsLevel1993,
  author =        {Rossignoli, R. and Ansari, A. and Ring, P.},
  journal =       {Phys. Rev. Lett.},
  month =         feb,
  number =        {8},
  pages =         {1061--1064},
  title =         {Projected Statistics and Level Densities},
  volume =        {70},
  year =          {1993},
  doi =           {10.1103/PhysRevLett.70.1061},
  issn =          {0031-9007},
}

@article{fantoParticlenumberProjectionFinitetemperature2017,
  author =        {Fanto, P. and Alhassid, Y. and Bertsch, G. F.},
  journal =       {Phys. Rev. C},
  month =         jul,
  number =        {1},
  pages =         {014305},
  title =         {Particle-Number Projection in the Finite-Temperature
                   Mean-Field Approximation},
  volume =        {96},
  year =          {2017},
  doi =           {10.1103/PhysRevC.96.014305},
  issn =          {2469-9985, 2469-9993},
}

@article{p.fantoProjectionVariationFinitetemperature2017,
  author =        {{P. Fanto}},
  journal =       {Phys. Rev. C},
  month =         nov,
  number =        {5},
  pages =         {051301},
  title =         {Projection after Variation in the Finite-Temperature
                   {{Hartree-Fock-Bogoliubov}} Approximation},
  volume =        {96},
  year =          {2017},
  doi =           {10.1103/physrevc.96.051301},
}

@article{zehSymmetryViolatingTrial1965,
  author =        {Zeh, H. D.},
  journal =       {Zeitschrift f{\"u}r Physik},
  month =         aug,
  number =        {4},
  pages =         {361--373},
  title =         {Symmetry Violating Trial Wave Functions},
  volume =        {188},
  year =          {1965},
  doi =           {10.1007/BF01326951},
  issn =          {0044-3328},
}

@article{robledoSignOverlapHartreeFockBogoliubov2009,
  author =        {Robledo, L. M.},
  journal =       {Phys. Rev. C},
  month =         feb,
  number =        {2},
  pages =         {021302},
  title =         {Sign of the Overlap of {{Hartree-Fock-Bogoliubov}}
                   Wave Functions},
  volume =        {79},
  year =          {2009},
  doi =           {10.1103/PhysRevC.79.021302},
  issn =          {0556-2813, 1089-490X},
}

@article{tanabeExtensionWickTheorem1999,
  author =        {Tanabe, K. and Enami, K. and Yoshinaga, N.},
  journal =       {Phys. Rev. C},
  month =         may,
  number =        {5},
  pages =         {2494--2499},
  title =         {Extension of {{Wick}}'s Theorem for Many-Particle
                   Matrix Elements},
  volume =        {59},
  year =          {1999},
  doi =           {10.1103/PhysRevC.59.2494},
  issn =          {0556-2813, 1089-490X},
}

@article{gambacurtaThermodynamicalPropertiesSmall2012,
  author =        {Gambacurta, Danilo and Lacroix, Denis},
  journal =       {Phys. Rev. C},
  month =         apr,
  number =        {4},
  pages =         {044321},
  title =         {Thermodynamical Properties of Small Superconductors
                   with a Fixed Number of Particles},
  volume =        {85},
  year =          {2012},
  doi =           {10.1103/PhysRevC.85.044321},
  issn =          {0556-2813, 1089-490X},
}

@article{bartelBetterParametrisationSkyrmelike1982,
  author =        {Bartel, J. and Quentin, P. and Brack, M. and Guet, C. and
                   H{\aa}kansson, H. -B.},
  journal =       {Nucl. Phys. A},
  month =         sep,
  number =        {1},
  pages =         {79--100},
  title =         {Towards a Better Parametrisation of {{Skyrme-like}}
                   Effective Forces: {{A}} Critical Study of the {{SkM}}
                   Force},
  volume =        {386},
  year =          {1982},
  doi =           {10.1016/0375-9474(82)90403-1},
  issn =          {0375-9474},
}

@article{chasmanDensitydependentDeltaInteractions1976,
  author =        {Chasman, R. R.},
  journal =       {Phys. Rev. C},
  month =         nov,
  number =        {5},
  pages =         {1935--1945},
  title =         {Density-Dependent Delta Interactions and Actinide
                   Pairing Matrix Elements},
  volume =        {14},
  year =          {1976},
  doi =           {10.1103/PhysRevC.14.1935},
  issn =          {0556-2813},
}

@article{reinhardAxialHartreeFock2021,
  author =        {Reinhard, P.-G. and Schuetrumpf, B. and
                   Maruhn, J. A.},
  journal =       {Comput. Phys. Commun.},
  month =         jan,
  pages =         {107603},
  title =         {The {{Axial Hartree}}\textendash{{Fock}} + {{BCS Code
                   SkyAx}}},
  volume =        {258},
  year =          {2021},
  doi =           {10.1016/j.cpc.2020.107603},
  issn =          {00104655},
}

@article{qiaoModelingSurvivalProbabilities2022,
  author =        {Qiao, C. Y. and Pei, J. C.},
  journal =       {Phys. Rev. C},
  month =         jul,
  number =        {1},
  pages =         {014608},
  title =         {Modeling Survival Probabilities of Superheavy Nuclei
                   at High Excitation Energies},
  volume =        {106},
  year =          {2022},
  doi =           {10.1103/PhysRevC.106.014608},
  issn =          {2469-9985, 2469-9993},
}

@article{goriely2008ImprovedMicroscopic,
  author =        {Goriely, S. and Hilaire, S. and Koning, A. J.},
  journal =       {Phys. Rev. C},
  month =         {Dec},
  pages =         {064307},
  publisher =     {American Physical Society},
  title =         {Improved microscopic nuclear level densities within
                   the Hartree-Fock-Bogoliubov plus combinatorial
                   method},
  volume =        {78},
  year =          {2008},
  doi =           {10.1103/PhysRevC.78.064307},
  url =           {https://link.aps.org/doi/10.1103/PhysRevC.78.064307},
}

@article{GORIELY2025139677,
  author =        {S. Goriely and S. Péru and S. Hilaire},
  journal =       {Phys. Lett. B},
  pages =         {139677},
  title =         {QRPA prediction of the nuclear level densities and
                   de-excitation photon strength functions},
  volume =        {868},
  year =          {2025},
  doi =           {https://doi.org/10.1016/j.physletb.2025.139677},
  issn =          {0370-2693},
  url =           {https://www.sciencedirect.com/science/article/pii/
                   S0370269325004381},
}

@article{zhangLevelDensityOddA2023,
  author =        {Zhang, Wei and Gao, Wei and Zhang, Gui Tao and
                   Li, Zhi Yuan},
  journal =       {Nucl. Sci. Tech.},
  month =         aug,
  number =        {8},
  pages =         {124},
  title =         {Level Density of Odd-{{A}} Nuclei at Saddle Point},
  volume =        {34},
  year =          {2023},
  doi =           {10.1007/s41365-023-01270-8},
  issn =          {1001-8042, 2210-3147},
}

@article{jiangNuclearLevelDensity2024,
  author =        {Jiang, X. F. and Wu, X. H. and Zhao, P. W. and
                   Meng, J.},
  journal =       {Phys. Lett. B},
  month =         feb,
  pages =         {138448},
  title =         {Nuclear Level Density from Relativistic Density
                   Functional Theory and Combinatorial Method},
  volume =        {849},
  year =          {2024},
  doi =           {10.1016/j.physletb.2024.138448},
  issn =          {03702693},
}

@article{wangNuclearLevelDensity2025,
  author =        {Wang, Jia Qi and Dutta, Saumi and Lv, Cui Juan and
                   Wang, Long Jun and Sun, Yang},
  chapter =       {Nuclear Structure},
  journal =       {Phys. Rev. C},
  month =         mar,
  number =        {3},
  pages =         {034324},
  publisher =     {{assocpub@aps.org}},
  title =         {Nuclear Level Density Studied in Odd-Mass Nuclei in
                   the Framework of the Projected Shell Model},
  volume =        {111},
  year =          {2025},
  doi =           {10.1103/PhysRevC.111.034324},
}

@article{chenShellmodelbasedInvestigationLevel2023,
  author =        {Chen, Jin Bei and Liu, Meng Lan and Yuan, Cen Xi and
                   Chen, Sheng Li and Shimizu, Noritaka and
                   Sun, Xiao Dong and Xu, Rui Rui and Tian, Yuan},
  journal =       {Phys. Rev. C},
  month =         may,
  number =        {5},
  pages =         {054306},
  title =         {Shell-Model-Based Investigation on Level Density of
                   {{Xe}} and {{Ba}} Isotopes},
  volume =        {107},
  year =          {2023},
  doi =           {10.1103/PhysRevC.107.054306},
  issn =          {2469-9985, 2469-9993},
}

@article{iljinovPhenomenologicalStatisticalAnalysis1992,
  author =        {Iljinov, A. S. and Mebel, M. V. and Bianchi, N. and
                   De Sanctis, E. and Guaraldo, C. and Lucherini, V. and
                   Muccifora, V. and Polli, E. and Reolon, A. R. and
                   Rossi, P.},
  journal =       {Nucl. Phys. A},
  month =         jul,
  number =        {3},
  pages =         {517--557},
  title =         {Phenomenological Statistical Analysis of Level
                   Densities, Decay Widths and Lifetimes of Excited
                   Nuclei},
  volume =        {543},
  year =          {1992},
  doi =           {10.1016/0375-9474(92)90278-R},
  issn =          {0375-9474},
}

@article{dilgLevelDensityParameters1973,
  author =        {Dilg, W. and Schantl, W. and Vonach, H. and Uhl, M.},
  journal =       {Nucl. Phys. A},
  month =         dec,
  number =        {2},
  pages =         {269--298},
  publisher =     {{Elsevier BV}},
  title =         {Level Density Parameters for the Back-Shifted Fermi
                   Gas Model in the Mass Range 40 {$<$} {{A}} {$<$} 250},
  volume =        {217},
  year =          {1973},
  doi =           {10.1016/0375-9474(73)90196-6},
  issn =          {0375-9474},
}

@article{fengProductionCrossSections2006,
  author =        {Feng, Zhao Qing and Jin, Gen Ming and Fu, Fen and
                   Li, Jun Qing},
  journal =       {Nucl. Phys. A},
  month =         may,
  pages =         {50--67},
  publisher =     {{Elsevier BV}},
  title =         {Production Cross Sections of Superheavy Nuclei Based
                   on Dinuclear System Model},
  volume =        {771},
  year =          {2006},
  doi =           {10.1016/j.nuclphysa.2006.03.002},
  issn =          {0375-9474},
}

@article{zubovSurvivalProbabilitySuperheavy2002,
  author =        {Zubov, A. S. and Adamian, G. G. and Antonenko, N. V. and
                   Ivanova, S. P. and Scheid, W.},
  journal =       {Phys. Rev. C},
  month =         jan,
  number =        {2},
  pages =         {024308},
  publisher =     {{American Physical Society}},
  title =         {Survival Probability of Superheavy Nuclei},
  volume =        {65},
  year =          {2002},
  doi =           {10.1103/PhysRevC.65.024308},
}

@article{xiaSystematicStudySurvival2011,
  author =        {Xia, Cheng Jun and Sun, Bao Xi and Zhao, En Guang and
                   Zhou, Shan Gui},
  journal =       {Sci. China Phys. Mech. Astron.},
  month =         aug,
  number =        {1},
  pages =         {109--113},
  title =         {Systematic Study of Survival Probability of Excited
                   Superheavy Nuclei},
  volume =        {54},
  year =          {2011},
  doi =           {10.1007/s11433-011-4438-2},
  issn =          {1869-1927},
}

@article{zhangPredictionsSynthesizingElements2024,
  author =        {Zhang, Ming Hao and Zhang, Yu Hai and Zou, Ying and
                   Wang, Chen and Zhu, Long and Zhang, Feng Shou},
  journal =       {Phys. Rev. C},
  number =        {1},
  pages =         {014622},
  publisher =     {{American Physical Society}},
  title =         {Predictions of Synthesizing Elements with {{Z}}=119
                   and 120 in Fusion Reactions},
  volume =        {109},
  year =          {2024},
  doi =           {10.1103/PhysRevC.109.014622},
  issn =          {2469-9985},
}

@article{ignatyukFissionPreactinideNuclei1975,
  author =        {Ignatyuk, A. V. and Itkis, M. G. and Okolovich, V. N. and
                   Smirenkin, G. N. and Tishin, A. S.},
  journal =       {Yad. Fiz.},
  month =         jun,
  number =        {6},
  pages =         {1185--1205},
  title =         {{Fission of pre-actinide nuclei. Excitation functions
                   for the ({$\alpha$},f) reaction}},
  volume =        {21},
  year =          {1975},
}

@article{dengExaminationPromisingReactions2023,
  author =        {Deng, Xiang Quan and Zhou, Shan Gui},
  journal =       {Phys. Rev. C},
  number =        {1},
  pages =         {014616},
  publisher =     {{American Physical Society}},
  title =         {Examination of Promising Reactions with {{Am}} 241
                   and {{Cm}} 244 Targets for the Synthesis of New
                   Superheavy Elements within the Dinuclear System Model
                   with a Dynamical Potential Energy Surface},
  volume =        {107},
  year =          {2023},
  doi =           {10.1103/PhysRevC.107.014616},
  issn =          {2469-9985},
}

@article{guttormsenConstanttemperatureLevelDensities2013,
  author =        {Guttormsen, M. and Jurado, B. and Wilson, J. N. and
                   Aiche, M. and Bernstein, L. A. and Ducasse, Q. and
                   Giacoppo, F. and G{\"o}rgen, A. and Gunsing, F. and
                   Hagen, T. W. and Larsen, A. C. and Lebois, M. and
                   Leniau, B. and Renstr{\o}m, T. and Rose, S. J. and
                   Siem, S. and Tornyi, T. and Tveten, G. M. and
                   Wiedeking, M.},
  journal =       {Phys. Rev. C},
  month =         aug,
  number =        {2},
  pages =         {024307},
  title =         {Constant-Temperature Level Densities in the
                   Quasicontinuum of {{Th}} and {{U}} Isotopes},
  volume =        {88},
  year =          {2013},
  doi =           {10.1103/PhysRevC.88.024307},
  issn =          {0556-2813, 1089-490X},
}

\end{document}